\documentstyle[12pt]{article}
\input epsf

\begin{document}
\begin{titlepage}
\thispagestyle{empty}
\title{
\vspace*{-1cm}
\begin{flushright}
{\small CPHT S 009.0201}
{\small KEK-TH-746}\\
\end{flushright}
\vspace{2.0cm}
Dispersion Relations  and Sum Rules for the  $\pi^0 \rightarrow \gamma
\gamma^*$ and
$\gamma^*
\rightarrow \gamma \pi^0$  Form Factor }
\vspace{2.0cm}
\author{Tran N. Truong \\
\small \em Centre de Physique Th\'eorique, 
{\footnote {unit\'e propre 014 du
CNRS}}\\ 
\small \em Ecole Polytechnique \\
\small \em F91128 Palaiseau, France \\
 and \\
\small \em Institute of Nuclear and Particle Studies, \\
\small \em High Energy Accelerator Research Organization,\\
\small \em 1-1 Oho, Tsukuba, Ibaraki, 305 - 0801 Japan } 

\date{February 2001}

\maketitle

\begin{abstract}
Sum rules for the off-shell isovector form factor $\pi^0 \rightarrow \gamma \gamma^*$ 
are given in terms of the
 pion form factor and the $\gamma \pi^{\pm} \rightarrow  \pi^{\pm} \pi^0$ experimental data.
Similarly, the corresponding sum rules for the off-shell isoscalar form factor are given in terms
of the experimental photon-3$\pi$ form factor and the $\gamma \pi \rightarrow 3\pi $ amplitude.
For some set of parameters, e.g $\Gamma (\rho \to \pi\gamma) \simeq 90 KeV$, the
$\pi^0
\rightarrow
\gamma
\gamma
$ rate and  slope parameter of the decay
$\pi^0
\rightarrow \gamma e^+e^-$ are in good agreement with the experimental data.
\end{abstract}
\end{titlepage}

Dispersion relation is an useful approach to calculate a physical amplitude when nonperturbative
effects are important. For example, in the presence of a low energy resonance, 
either one has to modify the perturbative results or to use dispersive approach in order to
satisfy the unitarity relation. The case of the vector pion form form factor, using dispersion
relation, was previously discussed \cite{truong1}. Sum rules for low energy parameters were
written down and were found to be in agreement with the experiment data. 

In this article we are interested in calculating the 
radiative electromagnetic form factor of the pseudoscalar meson $\pi^0$, for example the form
factor of 
$\pi^0$ decaying into
$\gamma e^+e^-$ or $e^+e^- \rightarrow \pi^0 \gamma$ using dispersion relation. These processes
are usually studied by the Vector Meson Dominance model \cite{sakurai} using the Gell-Mann, Sharp
Wagner (GMSW) model   \cite{wagner} which includes some  non perturbative effects.
We want to show that the dispersive approach enables us to express the calculated results in
terms of the measurable quantities with a better precision than the GMSW model. Some inacuracies
and difficulties of the vector meson dominance using the GMSW scheme can thus be avoided
\cite{rudaz, schecter, bando}.

  Our line of approach is similar to the study of the pion form factor \cite{truong1, truong2}.
The theoretical calculation  was based on the solution of the integral
equation of the  Muskellishvilli-Omnes  type \cite{mus, omnes}, using the experimental
information on the P-wave phase shifts in the low energy region up to 1 GeV. This approach yields
a r.m.s. radius of the pion too low by 10\% and the square of the magnitude of the pion form
factor  at the $\rho$ resonance (i.e. the leptonic width) by 30\% \cite{truong1, truong2}. This
is also the precision of the VMD model. The GMSW model \cite{wagner}, which used the VMD model
\cite{sakurai}, should therefore be accurate to this level. 

 On the
other hand, if the information on the P-wave phase shifts was combined with the experimental
magnitude of the time like pion form factor,  the r.m.s. of the pion  could be evaluated by a sum
rule which gave a result  in agreement with the data to  a few percents
\cite{truong1}. Likewise, if the solution of the Omnes-Muskellishvilli integral equation was used
with the input of the experimental value of the r.m.s. radius, the accuracy of the leptonic width
of the $\rho$ resonance could also  be calculated with an  accuracy of a few percents
\cite{truong1, truong3, truong4}.

In using dispersion relation for the 3-point function, one has the choice of dispersing or taking
either one the three particles involved off its mass shell, keeping the other two on their
mass-shell. For the case of $\pi^0
\rightarrow\gamma \gamma^*$ or $\gamma^* \to \gamma\pi^0$,  one disperses in one of  the photon
mass squared variable. In the following, we denote the dispersed photon as $\gamma^*$
 and the on shell photon 
simply as $\gamma$. 
 Unlike the GMSW scheme \cite{wagner}, we do not take two photons simultaneously off  their mass
shells, for which dispersion relations are not justified from a more fundamental viewpoint.

Proceeding with the GMSW scheme, one can calculate the $\pi^0 \to \gamma\gamma$ amplitude which
was found to agree with the $\pi^0 \gamma\gamma$ anomaly \cite{bell} and also with the
experimental data \cite{pdg}. In this calculation, one uses the experimental data on the leptonic
widths of
$\rho$,
$\omega$, the
$\omega
\to \pi\gamma$ width and the $SU(3)$ relation relating $\rho\pi\gamma$ vertex to
$\omega\pi\gamma$. The
$\gamma 3\pi$ amplitude
\cite{rudaz, schecter, bando} in the chiral limit  is a factor of $3/2$ too large compared  with
that given by the anomaly
\cite{adler}.  Although the introduction of a contact term in the $\gamma \to 3\pi$ amplitude
could solve this problem \cite{bando}, there is a problem with unitarity 
which will be discussed elsewhere \cite{truong5}.

It is therefore desirable to use dispersion relation, in combination with unitarity, 
 to solve this problem.

For the decay process $\pi^0 \rightarrow \gamma(k) \gamma^*(q)$, neglecting the electron
positron masses, the kinematics requires that $0\leq q^2\leq m_\pi^2$. 
The space-like radiative form factor of $\pi^0$  can be measured by the two photon processes in
$e^+e^-$ collision, $e^+e^- \rightarrow e^+e^- \pi^0$, keeping one of the  exchanged  photon
almost on its mass shell and the other photon off its mass shell. The limit of the space-like
momentum transfer is not limited by the $\pi^0$ mass, but by the kinematics of the two photon
process.
Likewise the time-like radiative form factor of the  $\pi^0 \gamma$ can be measured by the
process $e^+e^- \rightarrow \pi^0 \gamma$ i.e. the production of $\pi^0$ and $\gamma$ by the
single photon exchanged process. There is no kinematical limitation of the time-like momentum
transfer. We shall study this form factor as a function of the momentum transfer squared from
$-1 GeV^2< q^2 <1 GeV^2$.

\section {The Isovector Matrix Element  }

The electromagnetic hadronic current is decomposed into two parts, the
isovector and  isoscalar contributions. The lowest number of pion contributions are, respectively,
two and three pion states. These are the most important states because they are associated with,
respectively, the $\rho$ and $\omega$ vector mesons. Dispersing
in the photon mass one must take into the contributions of the 2$\pi$ and 3$\pi$ states with the
total angular momentum corresponding to the photon or $J=1^-$ states. Because of the G-Parity
invariance for strong interaction, these two states are orthogonal and we can therefore consider
their contribution separately. The final result for the form factors $\gamma^* \rightarrow \gamma
\pi$ or 
$\pi \to \gamma \gamma^* $ is
the algebaic sum of their contributions.

 The isovector and isoscalar contributions  are denoted,
respectively, as
$F^{3}(q^2)$ and $F^{0}(q^2)$. More explicitly the isovector contribution to the matrix element
 can be written as:
\begin{equation}
 M^3(\pi(p)) \rightarrow \gamma(k) \gamma^*(q))=-\epsilon^{\mu\nu\sigma\tau}e_\mu
k_\nu \epsilon_\sigma q_\tau F_{\gamma\gamma^*}^3(q^2) \label{eq:pgg*}
\end{equation}
and similarly the corresponding amplitude for the isoscalar contribution is denoted by
$F_{\gamma\gamma^*}^0(q^2)$. In the following, for convenience, we set $q^2 \equiv s$.

At zero momentum transfer $s=0$, the off-shell photon becomes the on-shell photon and
hence the matrix element in Eq. (\ref{eq:pgg*}) is just given by the matrix element of $\pi^0
\rightarrow
\gamma \gamma$ and is given by the chiral anomaly:
\begin{equation}
F_{\gamma\gamma}= \frac{e^2}{4\pi^2f_\pi} \label{eq:pgg}
\end{equation}
where the number of color $N_c=3$ has been used and $f_\pi=93 MeV$ is the pion decay constant and
$e$ is the electric charge. It is generally assumed that the matrix element for the physical
pseudoscalar
$\pi^0$ is not very much different from its chiral value, Eq. (\ref{eq:pgg}). This is a plausible
assumption, which will be used in this article, because it is well verified for $\pi^0 \rightarrow
2\gamma
$ decay.

It is clear that $F_{\gamma\gamma^*}^3(s)$ is an analytic function in the complex $s$ plane.
Depending on the assumption on its asymptotic behavior and the precision of the calculation, an
unsubtracted, once or twice subtracted  dispersion relation for 
$F_{\gamma\gamma^*}^3$ can be written. Similarly its first, second ... derivatives at any point
except on the positive real axis with $s\geq 4m_\pi^2$ can be written in terms of its imaginary
part. For simplicity in  writing  we shall drop, at the moment, the superscript and subscript of
the funcion
$F_{\gamma\gamma^*}^3$ and simply write it as $F(s)$.

Because the  form factor $F(s)$ is an analytic function with a cut from
$4m_\pi^2$ to
$\infty$, the $n^{th}$ times subtracted dispersion relation for $F(s)$ reads: 
\begin{equation}
F(s)=a_0+a_1s+...a_{n-1}s^{n-1}+ \frac{s^{n}}{\pi}\int_{4m_\pi^2}^\infty
\frac{ImF(z)dz}{z^{n}(z-s-i\epsilon)}
\label{eq:ff1} 
\end{equation}
where $n\geq 0$ and, for our purpose, the series around the origin is
considered. Because of the real analytic property of $F(s)$, it is real below $4m_\pi^2$. By
taking the real part of this equation, 
$ReF(s)$ is related to the principal part of the dispersion integral involving the $ImV(s)$
apart from the subtraction constants $a_n$.

 The
polynomial on the R.H.S. of Eq. (\ref{eq:ff1}) will be referred in the following as the
subtraction constants and the last term on the R.H.S. as the dispersion integral (DI). The
evaluation of DI as a funtion of $s$ will be done later. 
  Notice that
$a_n=F^n(0)/n!$ is the coefficient of the Taylor series expansion for
$F(s)$, where
$F^n(0)$ is the nth derivative of
$F(s)$ evaluated at the origin. 

 In theoretical work such as in perturbation
theory, the number of subtractions is minimal in the sense  that it just makes the DI converges.
In general, the more subtractions we make, the better is the suppression of the high energy
contribution which, in general,is difficult to calculate. This high energy suppression makes
the calculation more  reliable,  but at the same time, makes the calculation  less
predictive because of the introduction of extra parameters.

$ImF(s)$ can be evaluated using the unitarity relation. Instead of taking a complete set of
intermediate states in the unitarity relation, we truncate the summation and keep only the two
pion $P$ state. We shall show below that this approximation is sufficiently accurate to analyse
the data  for $s$ below 1 $GeV^2$.

It is straightforward to show in this case:
\begin{equation}
ImF(s) = \frac{e}{96\pi} \frac{(s-4m_\pi^2)^{3/2}}{s^{1/2}}V^*(s)G_1(s) \label{eq:unit}
\end{equation}
where $V(s)$ is the vector pion form factor with the normalisation $V(0)=1$ and $G_1(s)$ is the
P-wave projection of the invariant amplitude for $\gamma \pi^+ \rightarrow \pi^+\pi^0$:
\begin{equation}
M(\gamma(k)\pi^0(q_+) \rightarrow \pi^+(p_+)\pi^0(p_0)) =i\epsilon^{\mu\nu\sigma\tau}q_0^\nu
q_+^\sigma q_-^\tau G(s, t, u)
\label{eq:stu}
\end{equation}
where $\epsilon$ is the photon polarisation, $s=(k+q_+)^2$, $t=(p_+-q_+)^2$ and $u=(p_0-q_+)^2$.
Because all particles involved are on shell, one has $s+t+u=3m_\pi^2$. In the center of mass
system, in terms of the scattering angle $\theta$, we have:
\begin{eqnarray}
t&=&\frac{3m_\pi^2-s}{2}+\frac{1}{2}(s-m_\pi^2)\sqrt{1-4m_\pi^2/s}\cos\theta \nonumber \\
u&=&\frac{3m_\pi^2-s}{2}-\frac{1}{2}(s-m_\pi^2)\sqrt{1-4m_\pi^2/s}\cos\theta \label{eq:tu} 
\end{eqnarray}

 The partial wave expansion
for
$G(s,t,u)$ is given as follows \cite{martin, hannah}:
\begin{equation}
G(s,t,u) = \sum_{odd  l} G_l(s)P_l^{\prime}(\cos\theta) \label{eq:pw}
\end{equation}
where $\theta$ is the scattering angle and hence
\begin{equation}
G_1(s) = \frac{3}{8\pi}\int d\Omega \sin^2\theta G(s,t,u) \label{eq:kin}
\end{equation}
In terms of the function $G(s,t,u)$ the differential cross section for the process
  $\gamma \pi^{\pm}
\rightarrow \pi^{\pm}\pi^0$ is 
\begin{equation}
\frac{d\sigma}{d\cos\theta} =
\frac{1}{1024\pi}(s-m_\pi^2)\frac{(s-4m_\pi^2)^{3/2}}{s^{1/2}}\sin^2\theta\mid G(s,t,u)\mid^2
\label{eq:x}
\end{equation}
In Eq. (\ref{eq:unit}), the elastic unitarity relation for the vector pion form factor requires
that the phase of $V(s)$ is the phase of the P-wave $\pi\pi$ strong interaction $\delta(s)$.
Similarly, the phase of for the P-wave amplitude  $G_1(s)$ of the amplitude $\gamma \pi^+
\rightarrow
\pi^+\pi^0$ is also $\delta(s)$. The product $V(s)^{*}G(s)$ is therefore real and hence we can
rewrite Eq. (\ref{eq:unit}) as:
\begin{equation}
ImF_{\gamma\gamma^*}^3(s) = \frac{e}{96\pi} \frac{(s-4m_\pi^2)^{3/2}}{s^{1/2}}\mid V(s)\mid \mid
G_1(s)\mid
\label{eq:unitt}
\end{equation}
where we have restored the subscripts and superscripts. This equation is ambiguous to a plus or
minus sign. In the following we take the positive sign by convention. Hence $ImF(s)$ is  only
given in terms of the measurable quantities,
$\mid V(s)
\mid$ and
$\mid G_1(s) \mid $.

 There are excellent experimental data on the vector pion form
factor but accurate experimental data on $\gamma \pi^+ \rightarrow \pi^+\pi^0$ are not available.
To proceed, at this stage, one has to use some theoretical studies on this process which will be
discussed below.

Besides the 2$\pi$, there are higher mass states contributing to the unitarity relation for
$ImF_{\gamma\gamma^*}^3(s)$. For example, the
$K
\overline{K}$ or the 4$\pi$ states in the form of the $\omega\pi$ or the $\rho\pi\pi$ states etc. 
could become important at an energy above 1 $GeV$. As long as we are interested in the
phenomology below this energy region, it may be safe to take only into account of the lowest
intermediate states which, for the isovector contribution, is the interacting two pion state or 
the
$\rho$ vector meson, and for the isoscalar contribution, the interacting three pion states or the
$\omega$ vector meson.

If we want to find out how important is the 2$\pi$ contribution in the form of $\rho$ compared
with the higher intermediate states, we can write an unsubtracted dispersion relation for
$F_{\gamma\gamma^*}^3(s)$ and then evaluate this amplitude at $s=0$ to compare with the amplitude
of
$\pi^0
\rightarrow 2\gamma$. Needless to say, such an use of an unsubtracted dispersion relation can be
opened to criticisms because the higher intermediate states are not strongly suppressed; we can
only hope that, at low energy, their contributions are small compared with the 2$\pi$ and hence 
can be neglected. This is the spirit of the Vector Meson Dominance model in the more precised
language of the dispersion theory.

On the other hand if the $e^+e^-$ spectrum in the $\pi^0 \rightarrow \gamma e^+e^-$
process is to be calculated the higher intermediate states are then suppressed by a factor
$s^{-2}$ (see below) which makes  the result more reliable. The higher the derivative is at
$s=0$, the larger is the suppression of the high energy contribution.

We shall parametrize phenomenologically the pion form factor data below 1.1 $GeV$ by the
following  formula \cite{truong4} which agrees well with the experimental data  \cite{barkov,
aleph}:

      \begin{equation}
         V(s) =  \frac{1+0.16 s/s_R} {1 -s/s_R - {1\over
96\pi^2f_\pi^2}\{(s-4m_\pi^2)
 H_{\pi\pi}({s}) + {2s/3}\}}  \label{eq:vu1}
\end{equation}
where $f_\pi=0.093 GeV$, and $s_{R}=30.2 m_\pi^2$ which gives a good  fit to the modulus of
the pion form factor below 1.1 $GeV$ and the P-wave $\pi \pi $ phase shifts in this energy range.
$H_{\pi\pi}({s})$ is a well-known integral over the phase space factor:
\begin{equation}
   H_{\pi\pi}(s) = (2 - 2 \sqrt{s-4m_\pi^2\over s}log{{\sqrt{s}+\sqrt{s-4m_\pi^2}\over
2m_\pi}})+i\pi\sqrt{s-4m_\pi^2\over s} \label{eq:H}
\end{equation}

Neither good experimental data on  $\mid G(s,t,u)\mid$ nor its  P-wave amplitude $G_1(s)$ are
available. In fact, the  data on the width $\Gamma(\rho \to \pi\gamma)$  are in contradiction
\cite{capraro, dolinsky}. Once the experimental data on
$\mid G_1(s)\mid$ are available they can directly be used in Eq. (\ref{eq:unitt}) to calculate the
$\pi^0\gamma$ form factor.

For the time being, it is reasonable to use some theoretical models to parametrize $G_1(s)$. This
can be done by assuming the elastic unitarity relation and dispersion relation for this process. A
more complicated  singular integral equation  than that of the Muskelishvili-Omnes type is
obtained
\cite{truong5}. Given the experimental P-wave $\pi\pi$ phase shifts, its solution, which is
related to the
$\Gamma(\rho \to \pi\gamma)$ width, can be obtained by a numerical method and is a sensitive
function of the second derivative of
$G_1(s)$ with respect to $s$ at
$s=m_\pi^2$. This problem is related to the existence of the contact term in the photon-3$\pi$
coupling
\cite{rudaz, schecter, bando}.  Because $G_1(s)$ is an analytic function in the
$s$ variable with a unitarity right hand cut extending from 4$m_\pi^2$ to $\infty$, and a left
hand cut on the negative s axis. On the right cut this function, by unitarity, must have the phase
of the strong P-wave
$\pi\pi$ interaction
$\delta(s)$. The left hand cut is due to the t and u channel contributions. From  general grounds,
the P-wave amplitude can be written as a product of the two cuts, the right and left hand
cuts. For our purpose, the left hand cut contribution will be, phenomenologically, represented by
the product of a double pole and a term linear in $s$. We have  hence:

\begin{equation}
\frac{G_1(s)}{G_1^0} = V(s) c\frac{a^2}{(a+s)^2}(1-s/b)
\label{eq:fit2}
\end{equation}

where $c,a, b$ are real and $G_1^0$ is the chiral anomaly of $\gamma \rightarrow
3\pi$:

\begin{equation}
G_1^0 = \frac{e}{4\pi^2f_\pi^3} \label{eq:anomaly2}
\end{equation}

Eq. (\ref{eq:fit2}), with the real parameters $a,b,c$, is flexible enough to fit  the 
solution of the theoretical calculation. It also gives the correct phase for the P-wave amplitude
of the process $\gamma \pi^+ \to \pi^+ \pi^0$ which are  the P-wave
$\pi\pi$ phase shifts below 1.1 $GeV$. This parametrization is most accurate in the energy
range
$0.20 GeV^2 < s <0.8 GeV^2$ where its contribution to the following low energy  sum rules is most
important.

Instead of characterizing the results of the theoretical calculation as a function of the input
second derivative values at $s=m_\pi^2$, it is more physical to express them as a function
of the
$\Gamma(\rho \to \pi\gamma)$ width which is defined by the value of Eq. (\ref{eq:fit2}) compared
with its corresponding Breit-Wigner formula at $s=m_\rho^2$.

 In Table 1, we give the values of the parameters as a function
of the $\Gamma(\rho \to \pi\gamma)$ widths.

Rewriting Eq. (\ref{eq:unitt}) in the unit of $F_{\gamma\gamma}$ as given by Eq. (\ref{eq:pgg})
one has:
\begin{equation}
\frac{ImF_{\gamma\gamma^*}^3(s)}{F_{\gamma\gamma}}=\frac{1}{96\pi
f_\pi^2}\frac{(s-4m_\pi^2)^{3/2}}{\sqrt{s}} \mid V(s)\mid ^2 c \frac{a^2}{(a+s)^2}(1-s/b)
 \label{eq:isov}
\end{equation}
Now knowing  $ImF_{\gamma\gamma^*}^3(s)$ we can write down sum rules and dispersion relation.

Let us begin with testing the notion of the vector meson dominance, which can be translated into
the dispersion language, by assuming  an unsubtracted dispersion relation for $F(s)$. This allows
us  to find out how much  the contribution of $ImF_{\gamma\gamma^*}^3(s)$ is in the $\pi^0
\rightarrow
\gamma \gamma$ amplitude.

Writing an unsubtracted dispersion relation for $F_{\gamma\gamma^*}^3(s)$ and set $s=0$ we have:
\begin{equation}
F_{\gamma\gamma^*}^3(0)=\frac{1}{\pi}\int_{4m_\pi^2}^\infty ds
\frac{ImF_{\gamma\gamma^*}^3(s)}{s}
\label{eq:charge}
\end{equation}
The use of an unsubtracted dispersion relation requires that
$F_{\gamma\gamma^*}^3(s)\rightarrow 0$ as $s \rightarrow \infty$. It  is model dependent
because the high energy contribution to $ImF_{\gamma\gamma^*}^3$ is only suppressed by a factor of
$s$ in the sum rule, Eq. (\ref{eq:charge}).

With this reservation, using
$ImF_{\gamma\gamma^*}^3(s)$ as given by Eq. (\ref{eq:unitt}), with the pion form
factor $\mid V(s)\mid$ given by Eq. (\ref{eq:vu1}) and with the $\gamma\pi \to \pi\pi$ amplitude
 given in Table 1, the matrix element of the isovector 
$\pi^0\to \gamma\gamma*$ can be calculated as a function of the
$\Gamma(\rho \to\pi\gamma) width$. The results of our calculation are given on the  5th column of
the Table 1.

We now want to calculate the energy dependence or the pair $e^+e^-$ in $\pi^0 \rightarrow \gamma
e^+e^-$ at low energy. This can sufficiently be done by evaluating the first derivative of
$F_{\gamma\gamma^*}^3(s)$ at the origin: 
\begin{equation}
{F_{\gamma\gamma^*}^3}^{'}(0)= \frac{1}{\pi}\int_{4m_\pi^2}^\infty ds
\frac{ImF_{\gamma\gamma^*}^3(s)}{s^2}  \label{eq:der1}
\end{equation}
and if more precision is needed one could include higher derivatives. For example the second
derivative is given by:
\begin{equation}
{F_{\gamma\gamma^*}^3}^{''}(0)= \frac{2}{\pi}\int_{4m_\pi^2}^\infty ds
\frac{ImF_{\gamma\gamma^*}^3(s)}{s^3}  \label{eq:der2}
\end{equation}
The sum rule for the first derivative at $s=0$ of $F_{\gamma\gamma^*}^3$ is quite reliable
because the unknown high energy contribution is suppressed by the factor $1/s^2$ in Eq.
(\ref{eq:der1}). The  slope is usually expressed in terms of of the slope parameter
$\beta^{i}$ defined as
$\beta^{i} s/{m_\pi^{0}}^2$ with $i=0,3$ for, respectively, the isoscalar and isovector
contributions.

In table 1, the slope parameter for the isovector contribution is also given in the 5th column.
  The precision of the experimental data do not
allow us to compare  the second derivative of $F_{\gamma\gamma^*}^3$. We shall compare the values
of the matrix elements and slopes after our calculation of the isoscalar contribution.

In evaluating the RHS of Eq. (\ref{eq:charge}) or Eq (\ref{eq:der1}), the P-wave phase space
factor in Eq. (\ref{eq:unitt}) suppresses strongly the low energy contribution of the modulus of
the amplitude
$\gamma\pi
\rightarrow \pi^+\pi^-$. They are therefore very insensitive to the low energy behavior of this
amplitude. The well-known problem of reconciling Vector Meson Dominance and the
$\gamma 3\pi$ chiral anomaly is not a problem here. It will be discussed in a separate study
\cite{truong5}.

\section {The Isoscalar Matrix Element  }

Dispersing in the isoscalar photon mass, we have to take into account of the lowest intermediate
state in the unitarity relation, the 3$\pi$ states. The total isospin of the 3 pions is
$I=0$ and its total angular momentum is $1^-$ hence they are spacely completely antisymmetric. 
The elastic unitarity contribution to the isoscalar amplitude is:
$F_{\gamma\gamma^*}^0(s)$. 
\begin{equation}
ImF_{\gamma\gamma^*}^0(s) = <\gamma\mid 3\pi><3\pi \mid \gamma\pi> \label{eq:unitarity}
\end{equation}
Again, after integrating over the internal variables for the 3 pion system, we can establish the
phase theorem for
$<\gamma\mid 3\pi>$ as in the case of the 2 pion contribution to the isovector form factor. The
RHS of Eq. (\ref{eq:unitarity}) is therfore real and we can rewrite it as:
\begin{equation}
ImF_{\gamma\gamma^*}^0(s) = \mid<\gamma\mid 3\pi>\mid \mid<3\pi \mid \gamma\pi>\mid
\label{eq:unitarityy}
\end{equation}
Hence $ImF_{\gamma\gamma^*}^0(s)$, similar to the isovector case, can be directly obtained  from
experimental data. There are good data for the process $\gamma^* \rightarrow 3\pi$ which shows
a complete  dominance of the $\omega$ contribution. 

The process $\gamma\pi \rightarrow 3\pi$ has not been measured. However, the 3 pion resonance 
is such a narrow one that it is possible to
approximate it accurately by the $\omega$ meson. This is also true  for the photon 3
pion vertex. 

Defining the photon $\omega$ coupling as $e m_\omega^2/g_\omega$ \cite{wagner} with the photon
invariant mass squared equal to the $\omega$ mass the $g_\omega$ can be determined from its
leptonic width: $\Gamma (\omega \rightarrow e^+e^-) = (4\pi\alpha^2/3) (m_\omega /g_\omega^2)
$. The $\omega \rightarrow \pi^0\gamma$ width is given by: $\Gamma(\omega \rightarrow
\pi^0\gamma)=
g_{\omega\pi\gamma}^2 E_\gamma^3/(12\pi)$. Using the experimental values, $m_\omega=782 MeV$,
$\Gamma_\omega=8.41 \pm .09 MeV$ and the branching ratio $\Gamma_{\omega
\rightarrow\pi\gamma}/\Gamma_\omega=8.5 \pm.5\%$, one can  easily verify:
\begin{equation}
F_{\gamma\gamma}^0=0.50\pm0.02 F_{\gamma\gamma}(1+\frac{s}{m_\omega^2}) \label{eq:omega}
\end{equation}
hence the $\omega$ contribution is just half of the $\pi^0 \gamma\gamma$ anomaly, Eq.
(\ref{eq:pgg}). The imaginary part of the isoscalar form factor can be straightforwardly deduced
from this result:\begin{equation}
\frac{ImF_{\gamma\gamma^*}^0(s)}{F_{\gamma\gamma}}=(0.50\pm.02)\frac{m_\omega^3 \Gamma_\omega(s)}
{(m_\omega^2-s)^2+(m_\omega\Gamma_\omega(s))^2} \label{eq:sigo}
\end{equation}
where $\Gamma(s)$ is strongly dependent on $s$ due to the three pion phase space. We shall
neglect altogether the real part of the $\omega$ self energy correction.

\section {Sum of the isovector and isoscalar matrix elements}
Knowing the imaginary part of the form factor, Eqs. (\ref{eq:isov}, \ref{eq:sigo}) and using
dispersion relation, there is no difficulty in calculating the form factor in the interval 
 $-1 GeV^2<s<1 GeV^2$. The energy dependence of the $\pi^0 \to \gamma \gamma^*$, as will be shown
below, requires us  to take the same sign for the imaginary part of the isoscalar and isovector
contributions.

\subsection { $\pi^0 \to\gamma\gamma$ and $\pi^0 \to \gamma\gamma^*$}
Let us first examine the $\pi^0 \to\gamma\gamma$ and $\pi^0 \to \gamma\gamma^*$ amplitudes. The
observable matrix element $F_{\gamma\gamma^*}(s)$ is the sum of the isoscalar and the isovector
matrix elements. They cannot experimentally be separated from each other. We want to find out how
much the two and three pion intermediate states (in the form of
$\omega$ and
$\rho$) contributing to the $\pi^0 \rightarrow \gamma\gamma$ amplitude. Using Eq.
(\ref{eq:omega}) and the results given on the fifth column of the Table 1, the total matrix
elements are given in Table 2. They should be compared with experimental value
\cite{pdg,behrend}
\begin{equation}
F_{\gamma\gamma^*}(s) = 1+(0.032\pm.004)s/m_{\pi^0}^2 \label{eq:exp}
\end{equation}
Comparing this result with those given by Table 2 it is clear that the matrix element
corresponding to $\Gamma(\rho \to \pi\gamma)=92 KeV$ is favored. The experimental errors are,
however, large.

Because we have more faith in the slope sum rule than the unsubtracted dispersion relation due to
the suppression of the high energy contribution the
prediction on the $\Gamma(\rho \to \pi\gamma)$ width can reliably be made once the slope
is better measured. 

\subsection{Time-Like and Space-Like  $\pi^0 \gamma$ Form Factor}:

Using the imaginary part of the isovector and isoscalar contributions to the  form factor
$F^{\gamma\gamma^*}(s)$ as given by Eqs. (\ref{eq:isov}, \ref{eq:sigo}) we can calculate the
s-dependence of the form factor from $-1 GeV^2 < s < 1 GeV^2$ with reliability by the following
subtracted dispersion relation:
\begin{equation}
\frac{F_{\gamma\gamma^*}(s)}{F_{\gamma\gamma}}= 1+\frac{s}{\pi} \int_{4m_\pi^2}^\infty
\frac{ImF_{\gamma\gamma^*}(z)dz}{z(z-s-i\epsilon)}
\label{eq:fff}
\end{equation}
where $ImF_{\gamma\gamma^*}(s)$ is given by the sum of the RHS of Eqs. (\ref{eq:isov},
\ref{eq:sigo}). 

Because of the overwhelmed dominance of the $\omega$ contribution to the form factor
$F_{\gamma\gamma}(s)$ around the $\omega-\rho$ region, there is little hope to get physics out of
the
$\rho$ and
$\omega$ region unless the accuracy of the  cross section measurement can be improved to a few
percents. Likewise, an improvement in the  determination of the space like behavior of the
form factor will enable us also to determine accurately the $\rho\pi\gamma$ width.

\section{Conclusion and Acknowledgements}

The main point of this paper is to show the usefulness of the dispersion relation approach which
has not effectively been used in the literature for this problem. It allows us to relate various
experimental measurements without making  strong assumptions. The non perturbative effects due
to  strong interaction are straightforwardly handled by dispersion relation and unitarity. 

Part of this is done at the Institute for Nuclear and Particle Studies at the KEK High Energy
Accelerator Research Organization. The author would like to thank Professors M. Kobayashi
and H. Sugawara for hospitality.

\newpage
\begin{center}
\vspace{5mm}

\begin{tabular}{|c||c|c|c||c|} \hline
$\Gamma(\rho \to \pi\gamma)$ & $c$ & $a$ & $b$ &Isovector Matrix Element ($\pi \to
\gamma\gamma^*)$\\
\hline 60&1.01&80&80 & $0.40+0.0145s/m_{\pi^0}^2$\\ \hline
77&0.97&80&100 & $0.44+0.0154s/m_{\pi^0}^2$\\ \hline
92&1.06&50&$\infty$ & $0.49+0.0171 s/m_{\pi^0}^2$\\ \hline
110&1.04&60&$\infty$ & $0.55+0.0193 s/m_{\pi^0}^2$\\ \hline
\end{tabular}

\vspace{2mm}

Table 1:Parameters describing the P-wave amplitude for $\gamma \pi \to \pi \pi$ and the isovector
matrix element in the unit of the anomaly $F_{\gamma\gamma}$ Eq. (\ref{eq:pgg}); $a$ and $b$ are
in the units of $m_{\pi^+}^2=1$. $\Gamma(\rho \to \pi\gamma)$ are in the unit of KeV.
\end{center}

\bigskip
\begin{center}
\vspace{5mm}

\begin{tabular}{|c|c|} \hline
$\Gamma(\rho \to \pi\gamma)$  &Total Matrix Element ($\pi \to
\gamma\gamma^*)$\\
\hline 60& $(0.90\pm.036)+(0.0292\pm.0012)s/m_{\pi^0}^2$\\ \hline
77 & $(0.94\pm.037)+(0.030\pm.0012)s/m_{\pi^0}^2$\\ \hline
92& $(0.99\pm.04)+(0.032\pm.0013) s/m_{\pi^0}^2$\\ \hline
110 & $(1.05\pm.04)+(0.034\pm.0014) s/m_{\pi^0}^2$\\ \hline
\end{tabular}

\vspace{2mm}
Table 2: Total matrix element is the sum of isovector and isoscalar matrix elements in the unit
of the anomaly $F_{\gamma\gamma}$ Eq. (\ref{eq:pgg}). $\Gamma(\rho \to \pi\gamma)$ are in the
unit of KeV. The errors come from those of the isoscalar contribution.

\end{center}

\end{document}